\documentclass[pra,superscriptaddress,10pt]{revtex4-2}

\usepackage[utf8]{inputenc}
\usepackage{amsmath, amssymb, amsfonts,amsthm,amsopn,amscd}
\usepackage{dsfont}
\usepackage{graphicx}
\usepackage{longtable}
\usepackage{titletoc}
\usepackage[section]{placeins}

\usepackage{mathtools}

\usepackage{upgreek}
\usepackage{xcolor}
\usepackage{nameref}

\DeclareMathOperator{\tr}{tr}
\newcommand{\bra}[1]{\mathinner{\langle #1|}}
\newcommand{\ket}[1]{\mathinner{|#1\rangle}}

\newcommand{\ketbra}[2]{\mathinner{|#1\rangle\langle#2|}}

\newcommand{\ot}[0]{\otimes}
\newcommand{\one}[0]{\mathds{1}}

\newcommand{\PP}{\mathcal{P}}
\newcommand{\nn}{\nonumber}
\usepackage[breaklinks]{hyperref}
\hypersetup{
    colorlinks,
    linkcolor={red!40!black},
    citecolor={blue!50!black},
    urlcolor={blue!20!black}
}

\begin{document}

\title{Typicality of Steering for Two-qubit States}

\begin{abstract}

Phenomena that slip beyond the grasp of our classical intuition reveal uniquely quantum effects that deepen our understanding of the physical world and enable advances in information processing, particularly in quantum communication and computation. One such phenomenon is quantum steering, whereby measurements performed by one party influence the conditional states of another when the two share an entangled quantum system. If the observed correlations cannot be explained by a local hidden state model, the state is said to be steerable. In this work, we investigate the typicality of this behavior: given a generic two-qubit state and $m$ Haar-random projective measurements, what is the probability of observing steering? We derive analytical expressions for the steering probability $\mathcal{P}_S$ of Werner states in two- and three-setting scenarios, the latter restricted to coplanar projective measurements on the Bloch sphere. For larger numbers of settings and various random states ensembles, we perform numerical analyses showing that $\mathcal{P}_S$ increases systematically with the number of measurements and substantially exceeds the corresponding probabilities associated with Bell nonlocality. Our results demonstrate that random states with minimal environmental coupling exhibit a high probability of steering for finite $m$ and approach genuine typicality, $\mathcal{P}_S=100\%$, as the number of settings increases. We provide a detailed characterization of $\mathcal{P}_S$ across different state ensembles and specific families, including Bell-diagonal and Werner states, identifying those with the greatest non-classical potential and highlighting their relevance for protocols in which steering serves as a key resource.
\end{abstract}

\author{Gerard Angl\`es Munn\'e}
\affiliation{Institute of Theoretical Physics and Astrophysics, University of Gdańsk, 80-308 Gda\'nsk, Poland}

\author{Pawe{\l} Cie\'sli\'nski}
\affiliation{Centre for Quantum Technologies, National University of Singapore, Singapore 117543, Singapore}

\author{Tam\'as V\'ertesi}
\affiliation{HUN-REN Institute for Nuclear Research, P.O. Box 51, H-4001 Debrecen, Hungary}

\author{Wies{\l}aw Laskowski}
\affiliation{Institute of Theoretical Physics and Astrophysics, University of Gdańsk, 80-308 Gda\'nsk, Poland}

\maketitle

\section{Introduction}

Quantum mechanics exhibits phenomena that cannot be explained within the framework of local realistic theories. The first major challenge to such models was presented in the celebrated Einstein--Podolsky--Rosen (EPR) paradox~\cite{EPR1935}. Subsequent investigations of this paradox led to some of the most profound demonstrations of nonclassical behaviors, namely Bell nonlocality and quantum steering. In the case of Bell nonlocality, the correlations observed in quantum systems cannot be accounted for by any local hidden-variable theory~\cite{Brunner_2014}. Quantum steering, in contrast, demonstrates that hybrid models combining local hidden variables with quantum local hidden states (LHS) are likewise insufficient to reproduce all observed correlations~\cite{Uola_2020}. A distinctive feature of quantum steering is its intrinsic asymmetry, implying that at least one of the parties involved must be trusted. Notably, many quantum information protocols naturally operate within this framework, and the incompatibility of quantum correlations with LHS models has been shown to provide an advantage in several informational tasks. These include quantum key distribution~\cite{Branciard_2012}, subchannel discrimination~\cite{Piani_2015}, randomness generation~\cite{Law_2014}, and quantum teleportation~\cite{Reid_2013}. Steering is also closely connected to the problem of joint measurability, and a direct mapping between the two can be established~\cite{Tulio_2014,Uola_2014,Uola_2015}.

In this work, we investigate the typicality of quantum steering as a resource. Specifically, we ask: how likely is it to observe steering for generic two-qubit states subjected to random projective measurements? Similar questions have previously been addressed in the context of other forms of nonclassicality, such as Bell nonlocality~\cite{Liang_2010,Shadbolt_2012,Wallman_2012,Fonseca_2015,Senel_2015,de_Rosier_2017,Lipinska_2018,Barasinski_2019,Barasinski_2020,Yang_2020,de_Rosier_2020,Pandit_2022,Barasinski_2021,Cieslinski_2025} and generalized contextuality~\cite{Rossi_2025}. The probability of violating Bell inequalities has been studied extensively in scenarios involving fixed states as well as generic pure and mixed states. In all these works, measurements were chosen as Haar-random projective measurements, an approach that we also adopt here. Remarkably, it has been shown that for pure states, Bell nonlocality becomes certain as either the number of measurement settings or the local Hilbert-space dimension increases~\cite{de_Rosier_2017,Lipinska_2018}. Importantly, these results do not rely on any specific Bell inequality but instead encompass entire Bell scenarios. Here, we pursue an analogous investigation within the framework of quantum steering.

For two-qubit systems, this question has previously been examined for Werner states using fixed steering inequalities and either orthogonal random measurements~\cite{Maquedano_2024} or generic random measurements~\cite{Skrzypczyk_2014}. We extend these studies by considering random mixed two-qubit states and multi-setting random measurement scenarios without relying on fixed steering inequalities. Our contributions include analytical results for Werner states in the two-setting and in the three-setting scenarios, where projective measurements are sampled randomly from a single plane of the Bloch sphere in the latter. A related question was investigated in~\cite{Schumacher_2023}, where the typicality of steering in bipartite qudit systems was studied through the violation of a correlation-matrix-based inequality. In that work, informationally complete POVMs were employed, and the steering violation probability (termed as the Euclidean volume ratio) for Alice to Bob steering was estimated using methods developed in~\cite{Sauer_2021}. Here, using the semidefinite programming strategy from~\cite{Pusey_2013,Skrzypczyk_2014,SDP_review}, our numerical analysis focuses on scenarios involving $m$ random projective measurements and several state ensembles, again without reference to a fixed steering inequality. We present comprehensive numerical results demonstrating that even highly mixed states can exhibit substantial steering probabilities, which increase with the number of measurement settings. Moreover, these probabilities are significantly larger---often by orders of magnitude---than those associated with Bell nonlocality. This contrast highlights fundamental differences between the two forms of nonclassicality and provides new insight into the prevalence of quantum steering.

\section{Preliminaries}
\subsection{Quantum steering}
\label{sec:steering}

We consider a scenario in which two parties, Alice and Bob, share a two-qubit state $\varrho^{AB}$. Each party can perform $m$ local two-outcome projective measurements, denoted by $\{A_x\}_{x=1}^m$ and $\{B_y\}_{y=1}^m$, respectively. If Alice performs measurement $A_x$ and obtains outcome $a$, then, according to the state-update rule of quantum mechanics, Bob is left with the conditional unnormalized state $\varrho_{a|x}^B$. The collection of all such conditional states is referred to as a \emph{steering assemblage}. One may ask whether this assemblage necessarily arises from Alice's measurements on the shared quantum state or whether it can instead be modeled as a mixture of states drawn from a probability distribution over some hidden variable $\lambda$. Formally, this amounts to determining whether the assemblage admits the decomposition
\begin{equation}
\varrho_{a|x}^B = \int d\lambda p(\lambda) p(a|x,\lambda)\sigma_\lambda .
\end{equation}
This description constitutes a \emph{local hidden state} (LHS) model, representing a hybrid framework that combines general local hidden-variable theories with separable quantum correlations. If no such model can reproduce the observed assemblage, quantum steering is demonstrated, meaning that Alice's measurements genuinely influence Bob's conditional quantum states. We denote this situation by $A \rightarrow B$. The converse direction, as well as the possibility of steering in both directions, is defined analogously.

For a given number of measurements and outcomes, the existence of an LHS model can be determined using a finite assemblage~\cite{Pusey_2013, SDP_review}. 
Suppose Alice performs $m$ measurements with $\ell$ outcomes, i.e., $x=1,\dots,m$ and $a=1,\dots,\ell$. The hidden variable can then be chosen as $\lambda=(\lambda_1,\dots,\lambda_m)$, with $\lambda_i \in \{1,\dots,\ell\}$. In this case, the assemblage to be considered is constructed as~\cite{SDP_review}
\begin{align}\label{eq:an}
\varrho_{a|x}^B = \sum_\lambda D(a|x,\lambda)\,\sigma^B_\lambda ,
\end{align}
where
\begin{align}\label{eq:d}
    D(a|x,\lambda) =
    \begin{cases}
        1 & \text{if } a=\lambda_x , \\
        0 & \text{otherwise} .
    \end{cases}
\end{align}
Given a bipartite state $\varrho_{AB}$, if there exists a steering assemblage $\{\sigma^B_\lambda\}$ satisfying Eq.~\eqref{eq:an}, with $\varrho_{a|x} = \mathrm{Tr}_A[(E_{a|x} \otimes \one_B)\varrho_{AB}]$, then the post-measurement assemblage is said to be \emph{unsteerable} (i.e., it admits an LHS model). Here, $E_{a|x}$ denotes the positive-operator-valued measure (POVM) element corresponding to measurement $A_x$ and outcome $a$. If the post-measurement assemblage admits an LHS model for all measurement scenarios, including the limit $m\rightarrow\infty$, the state $\varrho_{AB}$ is called unsteerable. For finite $m$, this can be tested using the semidefinite program (SDP)~\cite{SDP_review}
\begin{align}\label{eq:SDP}
    \max_{\mu,\,\{\sigma^B_\lambda\}} \quad & \mu \\
    \text{s.t.} \quad & \sum_\lambda D(a|x,\lambda)\,\sigma^B_\lambda = \varrho_{a|x}^B, \nonumber \\
    & \sigma_\lambda^B - \mu \one \succeq 0 . \nonumber
\end{align}
If the optimal value satisfies $\mu\geq 0$, the assemblage admits an LHS model and is therefore unsteerable. Since steering is inherently asymmetric, a violation of this condition, i.e., $\mu<0$, implies that Alice can steer Bob's state, corresponding to the scenario $A\rightarrow B$. By interchanging the roles of Alice and Bob, one can similarly test for steering in the opposite direction. If steering is possible in both directions, we refer to the phenomenon as \emph{two-way steering}. Throughout this work, we distinguish the following cases:
\begin{enumerate}
\item[i)] $S_{AB}$: $A \rightarrow B$ and $B \rightarrow A$,
\item[ii)] $S_A$: $A \rightarrow B$ but not $B \rightarrow A$,
\item[iii)] $S_B$: $B \rightarrow A$ but not $A \rightarrow B$,
\item[iv)] $S_{\emptyset}$: neither $A \rightarrow B$ nor $B \rightarrow A$,
\end{enumerate}
and unless stated otherwise, we restrict our analysis to projective measurements.

\subsection{Steering probability}

Our main task is to investigate the typicality of quantum steering. A natural framework for addressing such question is provided by randomized measurements~\cite{Cieslinski2024,Elben_2022}. This approach has proven useful in a variety of quantum-information tasks, including entanglement detection and characterization~\cite{Tran2015,Tran2016,Brydges2019,Ketterer2019,Wyderka2020,Zhou2020,Elben2020,Yu2021,Imai_2021,Neven2021,Ketterer_2022}, state-function estimation~\cite{Brun2004,vanEnk2012,Knips2020}, quantum metrology~\cite{Rath_2021,Imai_2026}, and the characterization of topological phases~\cite{Elben2020topo}, among others. Moreover, all previously reported results concerning the probability and typicality of Bell inequality violations were obtained within this framework. Here, we seek to determine the probability of observing steering for a mixed two-qubit state $\varrho^{AB}$ subjected to $m$ Haar-random local projective measurements performed by either Alice or Bob. It is important to note that the detection of steering is conclusive for any finite number of measurement settings. In contrast, the failure to detect steering does not imply the existence of an LHS model, as additional measurements may still reveal steerability. Only in the asymptotic limit $m\rightarrow\infty$ can unsteerability be claimed. Nevertheless, for convenience, we will refer to states belonging to the class $S_{\emptyset}$ as unsteerable. Since all results are presented explicitly as functions of $m$, this terminology should not lead to ambiguity.

Unlike the pure states, where a preferred measure over the set of states exists (namely, the Haar measure), the notion of typicality for mixed states depends on the chosen ensemble.
The reason for focusing on mixed states is that up to a zero-measure set, all Haar-random pure states of two qubits are entangled, and all Haar-random projective measurements are incompatible. Therefore, observing steering in this scenario is certain. 
Consequently, the question of steering typicality becomes substantially more meaningful for mixed-state ensembles.  Throughout this work, the scenarios under consideration will be denoted by $\{\mathcal{E},m\}$, where $\mathcal{E}$ specifies the ensemble of quantum states and $m$ the number of measurement settings. Following~\cite{Fonseca_2015,Barasinski_2020,Barasinski_2021}, we define the steering probability as
\begin{equation}
    \mathcal{P}_S^{\{\mathcal{E},m\}} = \int d\Omega \, g(\Omega),
\end{equation}
where the integration is performed over all parameters defining a given steering scenario. The function $g(\Omega)$ is binary and outputs $0$ or $1$ depending on the existence or non-existence of an LHS model for the given parameters, respectively. Thus, $\mathcal{P}_S^{\{\mathcal{E},m\}}$ quantifies the probability of observing steering within the ensemble $\mathcal{E}$ for a fixed number of measurement settings $m$.

\subsection{Mixed-State ensembles}

The mixed-state ensembles we consider are defined via measures induced by the partial trace over a joint system of higher dimension. Let $|\Psi \rangle$ be a pure Haar-random state from the joint space $
{H}_{d \cdot k} = H^A \otimes H^B \otimes H^C,$ where $d = \mathrm{dim}(H^A \otimes H^B)$ is the dimension of the system of interest and $k$ is the dimension of the ancilla $C$. Tracing out the degrees of freedom associated with $C$ yields
\begin{equation}
    \varrho^{AB}_k = \mathrm{Tr}_C \big[|\Psi \rangle \langle \Psi | \big],
    \label{eq:reduced_k}
\end{equation}
which defines an element of the mixed-state ensemble $\mathcal{E}_k$. Sampling from $\mathcal{E}_k$ can be performed more efficiently via the Ginibre construction~\cite{Zyczkowski_2001}. Specifically, one generates a $d\times k$ Ginibre matrix $G_{d,k}$ whose complex entries are independently sampled from the normal distribution $\mathcal{N}(0,1)$, and constructs the density matrix as
\begin{align}
    \varrho^{AB}_k = \frac{G_{d,k} G_{d,k}^\dagger}{\mathrm{Tr}(G_{d,k} G_{d,k}^\dagger)} \, .
\end{align}
This procedure corresponds exactly to the sampling defined in Eq.~\eqref{eq:reduced_k}~\cite{Zyczkowski_2001}. For $k=1$, all states are pure, while for $k=d$ the distribution coincides with the Hilbert–Schmidt distribution. Hence, we denote $\mathcal{E}_d = \mathcal{H}$.

Many properties of states from $\mathcal{E}_k$ are already known. For example, the average purity of states within it is given by~\cite{Zyczkowski_2001}
\begin{equation}
    \langle \mathrm{Tr}(\varrho^2) \rangle_{d,k} = \frac{d+k}{dk+1}.
\end{equation}
Even more interestingly, the probability of sampling an entangled two-qubit state from this ensemble is $1$, $13/14 \approx 0.928$, and $25/33 \approx 0.757$ for $k=2,3,4$, respectively~\cite{Slater_2015}. These results will be used later in the discussion of the steering probabilities $P_S^{\{\mathcal{E},m\}}$.

\section{Steering Probability Results}

We begin by presenting our numerical findings for the steering probability, considering up to $m=10$ measurement settings and mixed-state ensemble $\mathcal{E}_k$ with $k=2,3,4$. We then turn our attention to more specialized cases. We examine $\mathcal{P}_S$ for Bell-diagonal states and evaluate the asymptotic regime $m \rightarrow \infty$. Next, we derive explicit analytical expressions for $\mathcal{P}_S$ in the case of two-qubit Werner states, with $m=2$ random measurement settings, as well as $m=3$ settings restricted to a single plane of the Bloch sphere. Finally, we investigate pure random states subjected to noise, specifically depolarizing channels and one-site separable noise, to assess the robustness of steering probability and its dependence on noise parameters.

\subsection{Generic multi-setting scenario}

The steering probability for the scenarios $\{\mathcal{E}_k, m\}$ was estimated by running the SDP in Eq.~\eqref{eq:SDP} on samples of $10^6$ (as denoted in Table~\ref{tab:P_S}) random two-qubit states from $\mathcal{E}_k$ and independently drawn $m$-tuples of Haar-random projective measurement settings. The latter were generated explicitly as $M = U_H |0\rangle \langle 0| U_H^{\dagger} - U_H |1\rangle \langle 1| U_H^{\dagger}$, where $|0\rangle$ and $|1\rangle$ are the $\pm 1$ eigenstates of the Pauli matrix $\sigma_z$, and $U_H$ is an $SU(2)$ matrix drawn according to the Haar measure.
The SDP calculations were performed using Mosek~\cite{mosek} through JuMP~\cite{jump}, with primal and dual feasibility tolerances set to $10^{-8}$.
The effective statistical uncertainty, including numerical and sampling errors, is estimated to be in the order of $10^{-3}$.

\begin{figure}[htbp]
    \centering
    \includegraphics[width=\linewidth]{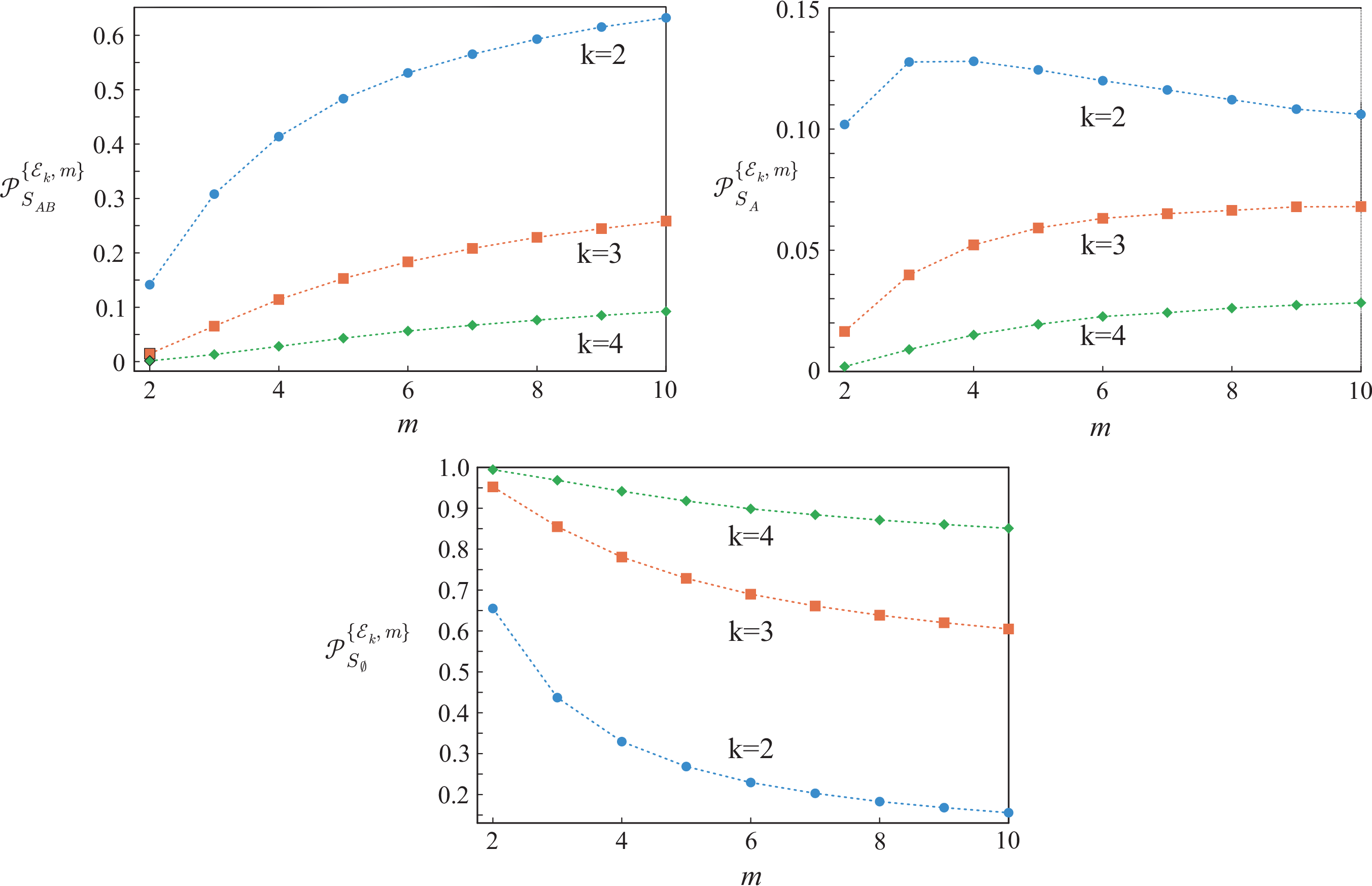}
    \caption{\textit{Steering probability for $S_{AB}$, $S_{A}$, and $S_{\emptyset}$ as a function of the number of measurement settings}. Each color corresponds to a different random state ensemble $\mathcal{E}_k$. Each data point is obtained by averaging the single-sample SDP output (see main text for details) over $10^6$ independent realizations.
    For the Hilbert--Schmidt ensemble ($k=4$), steering is almost negligible for $S_{AB}$ and $S_A$ at $m=2$, but increases to approximately $10\%$ and $7.5\%$, respectively, at $m=10$. For $k=3$, the steering probabilities grow rapidly with the number of measurements, reaching nearly $33\%$ for the $S_{AB}$ scenario.
    The most pronounced effects are observed for the minimal coupling case of $k=2$, where steering becomes the dominant feature of the ensemble for $m > 2$. At $m=10$, the probability for $S_{AB}$ reaches approximately $65\%$, while $S_{\emptyset}$ decreases rapidly to about $15\%$. In the limit $m \rightarrow \infty$, the latter approaches $0\%$. For the $S_A$ case, the probability is non-monotonic: it starts at approximately $10\%$, reaches a maximum at $m=4$, and returns to about $10\%$ at $m=10$. Further discussion is provided in the main text.}
    \label{fig:P_S_one-way}
\end{figure}

Since the considered ensembles are not necessarily symmetric, we compute steering probabilities for the scenarios $S_{AB}$, $S_A$, $S_B$, and $S_{\emptyset}$, as defined in Sec.~\ref{sec:steering}. This allows us to resolve the structure of steerable correlations and the distribution of one- and two-way steering. The corresponding results are shown in Figs.~\ref{fig:P_S_one-way}. In all cases, states drawn from $\mathcal{E}_k$ exhibit symmetry under exchange of parties, i.e., $\mathcal{P}_S^{\{\mathcal{E}_k,m\}}$ for $S_A$ and $S_B$ agree within the method's estimated uncertainty. This symmetry can be understood via the filtering argument presented later in Sec.~\ref{sect:werner} (see Eq.~\eqref{eq:filter}), and therefore only one of the two is shown.

For $k=4$, corresponding to the Hilbert--Schmidt ensemble $\mathcal{E}_4=\mathcal{H}$, the steering probability for $m=2$ measurements per party is approximately zero for both $S_{AB}$ and $S_A$. Consistently, the probability of observing no steering ($S_{\emptyset}$) is close to $100\%$. Increasing the number of settings leads to a clear improvement, with $\mathcal{P}_S^{\{\mathcal{H},10\}}$ reaching approximately 
$10\%$, $4\%$, and $85\%$ for $S_{AB}$, $S_A$ (or $S_B$), and $S_{\emptyset}$, respectively. Given that the fraction of entangled states in this ensemble is $75.7\%$, further increases in $m$ are expected to improve the observed steering typicality.

The situation changes significantly for $k=3$. While the initial steering probabilities for $S_{AB}$ and $S_A$ are small, they increase rapidly with the number of measurements, reaching approximately $33\%$ for $S_{AB}$ and $7.5\%$ for $S_A$ at $m=10$. At the same time, the probability of observing no steering decreases from about $95\%$ to approximately $60\%$.

As expected, ensembles with a larger proportion of entangled states exhibit more typical steering behavior. For $\mathcal{E}_2$, the probability of observing $S_{AB}$ reaches $63.2\%$ at $m=10$, while $S_{\emptyset}$ decreases rapidly to about $15\%$. In this case, the asymptotic behavior of $S_{\emptyset}$ can be determined exactly. From Schmidt decomposition, up to a zero-measure set, all $k=2$ states are rank-2. Based on the results from~\cite{Zhang_2025}, all such states are steerable and hence $\mathcal{P}_{S_{\emptyset}}^{\{\mathcal{E}_2,m \rightarrow \infty \}}=0\%$. Consequently, at least one kind of steering is always present in the asymptotic limit.
Interestingly, for the $S_{A}$ scenario, the probability of interest is around $10\%$ for $m=2$. Then, it slightly increases until it reaches a maximum in $m=4$, and decreases to its approximate initial value for $m=10$.
This behavior can be explained as follows: as expected, the steering probability for $S_{A}$ initially increases with $m$. However, as $m$ continues to increase, it becomes more likely that if $A$ steers $B$, $B$ will also steer $A$, which reduces the probability that $S_{A}$ occurs. This increasing and then decreasing leads to the observed maximum.

From these findings, it is clear that across all examined ensembles, steering occurs in a significant fraction of two-qubit states. Given a sufficiently large number of random projective measurement settings, steering can be regarded as typical. In particular, with the minimal coupling to the environment ($k=2$), steering becomes a dominant feature of the ensemble for $m>2$, i.e., the probability for $S_{\emptyset}$ is $<50\%$. Asymptotically, as discussed before, the steering becomes truly typical.
Nevertheless, for other $k$ and the minimal operational case of $m=2$, steering for $S_{AB}$ and $S_{A}$ is almost non-existent.
While specifically tailored measurements could improve their ability to exhibit non-LHS behavior, in the generic case, this phenomenon is negligible.

One may also ask how these values of $\mathcal{P}_S$ compare to the probability of Bell inequalities violation. Using techniques first introduced in~\cite{de_Rosier_2017}, we evaluate this probability for $2 \leq  m \leq 10$ within the Hilbert--Schmidt ensemble $\mathcal{H}$ on a sample size between $10^5$--$10^9$. The corresponding results are reported in Table~\ref{tab:P_S}. It is evident that the steering probability, including in the one-way scenario, is at least one order of magnitude larger than the probability of violating a Bell inequality. While this qualitative difference is expected, the quantitative gap is surprisingly large.

\begin{table}[h!]
\begin{tabular}{c|ccccc} \hline
 & \multicolumn{3}{c}{$\mathcal{P}_S^{\{\mathcal{H}, m \}}[\%]$} \\ \hline
$m$ & Nonlocality & $S_A \,(S_B)$ & $S_{AB}$ & $1-S_{\emptyset}$ \\ \hline
2  & 0.001 ($10^9$) & 0.202  &  0.141 & 0.550  \\
3  & 0.007 ($10^9$) & 0.909  & 1.313 & 3.152 \\
4  & 0.020 ($10^9$) & 1.508 & 2.809  & 5.839  \\
5  & 0.046 ($10^6$) & 1.945  & 4.316 & 8.219  \\
6  & 0.062 ($10^6$) & 2.268 & 5.632  &  10.139  \\
7  & 0.100 ($10^6$) & 2.428 & 6.692 &  11.590 \\
8  & 0.128  ($10^6$) & 2.616  & 7.636 & 12.882 \\
9  & 0.156 ($10^5$)&  2.741  & 8.502 & 13.955 \\
10 & 0.197  ($10^5$) &2.833  & 9.233 & 14.898 \\ \hline
\end{tabular}
\caption{\label{tab:P_S} Steering probabilities $P^{\{\mathcal{H},m\}}_S$ for the Hilbert-Schmidt ensemble of random two-qubit states with $m$ Haar random projective measurements compared with the Bell-inequality violation probability in the corresponding scenario. The sample size used to estimate the probability for Bell nonlocality is given explicitly in parentheses. For the steering probability, it was set to $10^6$. Different steering scenarios $S$ correspond to the four examined cases defined at the end of Sec.~\ref{sec:steering}. The above data shows that the typicality of steering for the examined ensemble is at least one order of magnitude greater than for the Bell nonlocality case. For comparison, the probability of drawing an entangled state from $\mathcal{H}$ is $25/33 \approx 75.7\%$.}
\end{table}

\subsection{Bell-diagonal states}

\begin{figure}[htbp]
    \centering
    \includegraphics[width=0.55\linewidth]{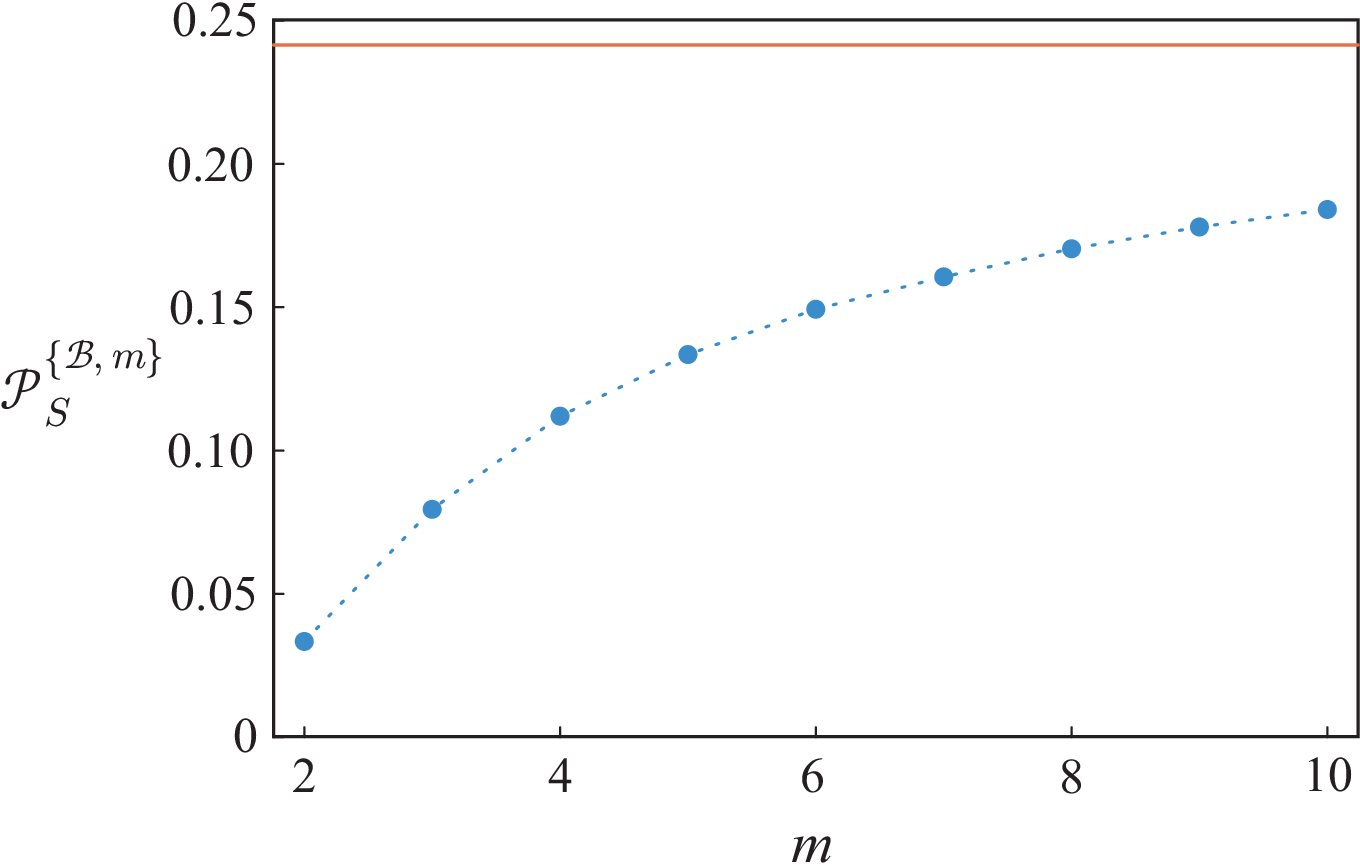}
    \caption{Steering probability $\mathcal{P}_S^{\{\mathcal{B},m\}}$ for random Bell-diagonal states as a function of $m$ Haar-random projective measurement settings. Initially, the steering probability is of the order of few percent. As expected, increasing $m$ causes it to grow, and reach almost $20\%$ for $m=10$. In the limit $m\rightarrow\infty$, the steering probability approaches $24.15\%$, indicated by the horizontal line.}
    \label{fig:P_S_Bell}
\end{figure}

Now we turn our attention to a more specialized class of states, namely Bell-diagonal states, defined as convex mixtures of the two-qubit Bell states. Explicitly,
\begin{equation}
    \varrho_{\mathcal{B}}=\sum^{4}_{i=1} p_i |\psi_i \rangle \langle \psi_i|,
    \label{eq:Bell_diag}
\end{equation}
where 
\begin{align}
 \ket{\psi_{1/2}}=\ket{\phi^{\pm}}&=(1/\sqrt{2}) (\ket{00}\pm \ket{11}), \nn \\
 \ket{\psi_{3/4}}=\ket{\psi^\pm}&=(1/\sqrt{2}) (\ket{10}\pm \ket{01})\,.
\end{align}
Random Bell-diagonal states are sampled from the ensemble $\mathcal{B}$ by drawing the coefficients  $\{p_i\}_{i=1}^4$ uniformly from the three-dimensional simplex.
Since the state is symmetric, the probability that $A$ steers $B$ is the same as the probability that $B$ steers $A$. Thus, it follows that $\PP^{\{\mathcal{B},m\}}_{S_A}=\PP^{\{\mathcal{B},m\}}_{S_B}=0$ and $\PP_{S_{AB}}^{\{\mathcal{B},m\}}=1-\PP_{\emptyset}^{\{\mathcal{B},m\}}$. As a consequence, we only need to compute one direction of steering probability $\PP^{\{\mathcal{B},m\}}_{S}$, which is the same as the two-way steering probability.
The results obtained using $10^6$ samples, and the same procedure as in the previous section, are shown in Fig.~\ref{fig:P_S_Bell}.

Interestingly, for this family, we can also evaluate the steering probability in the limit of infinitely many measurements. In the asymptotic case $m\rightarrow\infty$, we use the construction from~\cite{Nguyen_2020} to estimate the volume of steerable states as $24.15\%$. Despite containing a smaller fraction of entangled states than the Hilbert--Schmidt ensemble $\mathcal{H}$, this ensemble exhibits a higher steering probability for all finite $m$ considered. This behavior, however, does not persist when compared to $\mathcal{E}_3$, where the steering probabilities are comparable but typically larger than those for random Bell-diagonal states.

\subsection{Werner state and random pure states under noisy channels}\label{sect:werner}

Here, we will extend on the results presented in~\cite{Maquedano_2024, Skrzypczyk_2014}, and provide an analytical derivation of $A \rightarrow B$ steering probability for the two-qubit Werner state 
\begin{equation}
\varrho_W(v) =v \ket{\psi^-}\bra{\psi^-}+\frac{1-v}{4} \one_4 \,\,, \label{rhoW}
\end{equation}
under Haar random projective measurements. As in the Bell-diagonal case, the state is symmetric, and thus the steering direction does not matter, i.e., the steering probability for $A \rightarrow B$ is the same as for $B \rightarrow A$, and we denote it by $\PP^{\{\mathcal{W},m\}}_S$. For clarity, we state that the parameter $v$ is not sampled in any way.

In the two-setting case, using the steering criterion from~\cite{Grinko_2025} and symmetry of Haar measure we show that
\begin{align}
\label{PSv}
\mathcal{P}_S^{\{\mathcal{W},2\}}(v)=\begin{cases} 0 \quad  &\text{for} \quad  0 < v < 1/\sqrt{2}\,,\\
\sqrt{2v^2-1}/v^2 \quad &\text{for} \quad 1/\sqrt{2} < v \leq 1\,.\end{cases}
\end{align}
See Appendix~\ref{AppA} for its derivation. Importantly,  the above constraints on the possible values of $v$ reproduce the Werner state critical visibility of $v_{crit}=1/\sqrt{2}$. From the symmetry of the state $B \rightarrow A$ steering probability follows trivially.

Extending the steering scenario to $m\leq 3$ settings we were able to find the steering probability for the situation in which the measurements are drawn randomly and uniformly from a fixed plane on the Bloch sphere. 
To make the difference clear, we denote the number of in-plane settings by $2'$ and $3'$ instead of $2$ and $3$.
In Appendix~\ref{AppB} we show that for $m=2'$ and $v \geq 1/\sqrt 2$, the steering probability is given as
\begin{equation}\label{eq:werequator2}
\mathcal{P}_S^{\{W,2'\}}(v) = \frac{4}{\pi}\arccos\left(\frac{1}{\sqrt{2}\,v}\right), 
\end{equation}
with $\mathcal{P}_S^{\{W,2'\}}(v)=0$, otherwise.
For the case of $m=3$ and $v \geq 2/3$, it can be expressed through the following integral
\begin{equation}\label{eq:werequator3}
\mathcal{P}_S^{\{W,3'\}}(v)  = \frac{32}{\pi^2} \int_{0}^{\pi/4} \min\Big(t, \arccos\frac{(1/v) - \cos(2t)}{2 \sin t}\Big) \, dt.  
\end{equation}
with $\mathcal{P}_S^{\{W,3'\}}(v) =0$ for $v < 2/3$.

\begin{figure}[h!]
    \centering
    \includegraphics[width=0.6\linewidth]{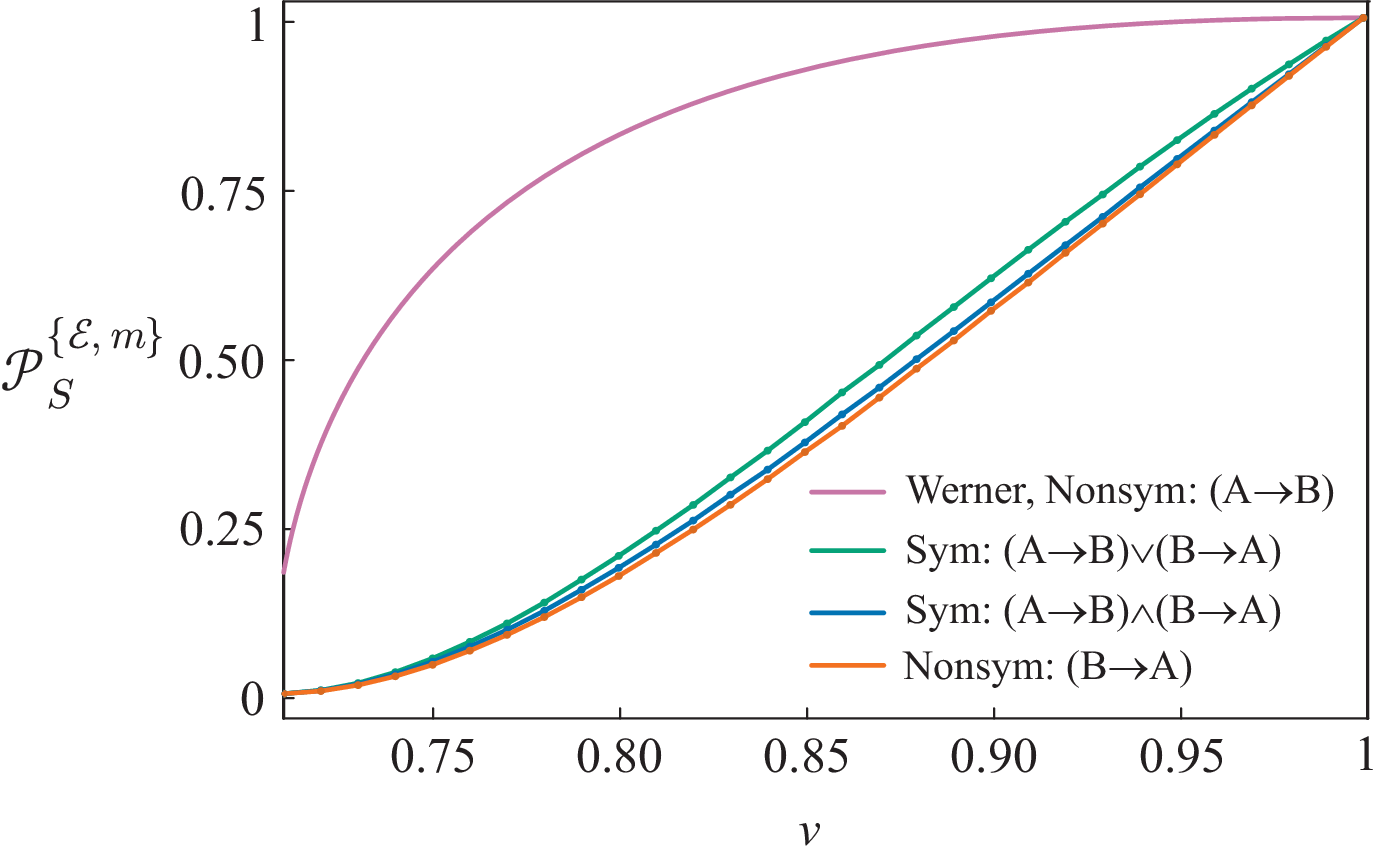}
    \caption{\textit{Steering probability for depolarized states as a function of visibility for $m=2$ Haar-random projective measurement settings.} The steering probability $\mathcal{P}_S^{{\mathcal{D}_v,m}}$ for Haar-random depolarized two-qubit states is shown in green and blue and denoted as Sym. The corresponding probability for Werner states is plotted in green. The orange curve represents the probability of $B \rightarrow A$ steering for Haar-random states subjected to local separable noise (denoted as Nonsym), as defined in Eq.~\eqref{eq:nonsym}.}
    \label{fig:werner}
\end{figure}

Motivated by the structure of the Werner state, we proceed to determining ${\mathcal P}_S$ for two-qubit Haar random depolarized states 
\begin{align}
\varrho =v\ketbra{\psi_H}{\psi_H}+\frac{1-v}{4} \one_4 \,.
\end{align}
where $\ket{\psi_H}$ is a two-qubit Haar random state. The set of those states we denote as $\mathcal{D}_v$. Since such states are not necessarily symmetric, we compute all possible steering probabilities, i.e.,  $A \rightarrow B$ ($B \rightarrow A$), two-way $A \rightarrow B \land A \rightarrow B $, and either-way $A \rightarrow B \lor A \rightarrow B$ steering. Here we denote all steering options explicitly to maintain clarity.
For each $v\in [0.7,1]$ and a step of $\Delta v=0.01$ we run the SDP~\eqref{eq:SDP} for $10^6$ times. Below that threshold, all evaluations yield 0. 
This indicates that the critical visibility for the two-qubit Haar random depolarized states is similar to the one obtained for the Werner state.
Our numerical results show that the steering probability $\mathcal{P}_S^{\{\mathcal{D}_v,m\}}$ does not depend on the direction, i.e., $\mathcal{P}_S^{\{\mathcal{D}_v,m\}}$ for $A\rightarrow B$ and for $B\rightarrow A$ is equal within the estimated error. 
This is expected since we are sampling states uniformly in all possible directions. 
However, not all of the sampled states are two-way steerable, which results in a difference between the steering probability for $A \rightarrow B \land A \rightarrow B $ and for $A \rightarrow B \lor A \rightarrow B$.
More specifically, the difference reaches its maximum when $v=0.9$, in which the probability of finding states that are not two-way steerable from the whole set that does not admit an LHS model is $\approx 5.8\%$.
Thus, most of the steerable states from $\mathcal{D}_v$ are two-way steerable.
The $\mathcal{P}_S^{\{\mathcal{D}_v,m\}}$ for $A \rightarrow B \land A \rightarrow B $ and $A \rightarrow B \lor A \rightarrow B$ is plotted in Fig.~\ref{fig:werner} (labeled as Sym), where one can see that the differences between them are small. 
Furthermore, the graphic shows that the steering probability for Haar random depolarized states is always below that of a Werner state with the same visibility.
This behavior is expected since Werner states are generally more entangled. 

Since $\mathcal{P}_S$ for Haar random depolarized states is independent of the steering direction, we propose examining the following state
\begin{align}\label{eq:nonsym}
\varrho^{AB}(v) =v\ketbra{\psi_H}{\psi_H}+\frac{1-v}{2} \left(\one_2 \ot \tr_A \left(\ketbra{\psi_H}{\psi_H}\right)\right)  \,.
\end{align}
where $\tr_A( \cdot )$ is the partial trace over $A$. This construction is analogous to the family of one-way steerable states studied previously, for example, in~\cite{Bowels_2016,Nguyen_2019}. Due to the asymmetry between the two subsystems, one may expect different steering probabilities for the $A\rightarrow B$ and $B\rightarrow A$ scenarios. 

Interestingly, for the $A \rightarrow B$ scenario, the obtained results followed Eq.~\eqref{PSv} exactly. Changing an isotropic noise to a separable one makes Haar random states equivalent to the Werner state in the steering probability sense.
This behavior can be understood based on the following filtering and locality arguments.  
Given the state $\varrho^{AB}$ in Eq.~\eqref{eq:nonsym}, we show that the assemblage 
\begin{equation}
\varrho_{a|x} = \tr_A[(E_{a|x} \otimes \one_2)\varrho^{AB}(v)],    
\end{equation}
for a given set of measurements $\{A_{x}\}$, is steerable from Alice to Bob if and only if the assemblage
\begin{equation}
\tilde\varrho_{a|x} = \tr_A[(E_{a|x} \otimes \one_2)\varrho_W(v)]    
\end{equation}
is steerable for the same set of measurements $\{A_{x}\}$. 
This result follows directly from two facts: (i) local filtering on Bob’s side does not change the steerability of the state, and (ii) 
\begin{equation}\label{eq:filter}
\varrho_W(v)=\frac{1}{\mathcal{N}}(\one_2\otimes F_B)\varrho_{AB}(v)(\one_2\otimes F_B)    
\end{equation}
holds for any non-product $\ket{\psi_H}$, where $F_B$ is an appropriate invertible local filtering operation and $\mathcal{N}$ denotes a potential renormalization factor. Fact (i) is taken from~\cite{Tulio_2015}, whereas fact (ii) is proven in Appendix~\ref{AppC} and follows a similar reasoning to the one from \cite{Bowels_2016}.

For $B \rightarrow A$ case, we repeat our routine for $v\in [0,1]$.
While for $v\in [0,0.7]$ the probability of steering is still 0, for $v\in [0.7, 1]$ it is slightly below the result obtained for Haar random depolarised states (see Fig.~\ref{fig:werner}). 
Furthermore, we also compute the two-way steering probability  $(A \rightarrow B \land B \rightarrow A )$ and the either-way one $(A \rightarrow B \lor A \rightarrow B )$.
Numerics show that the probability for two-way steering is equal to the one for $(B \rightarrow A)$ and the probability for either-way steering is equal to the one for $A \rightarrow B$. Thus, for the non-symmetric state with separable noise defined in Eq.~\eqref{eq:nonsym}, we conclude that, if $B \rightarrow A$ then $A \rightarrow B$, however the converse does not hold.

\section{Conclusions}

In this work, we investigated the typicality of quantum steering for generic two-qubit states subjected to Haar-random projective measurements. We derived analytical expressions for the steering probability of Werner states in both two-setting and three-setting scenarios, the latter restricted to projective measurements confined to a single plane of the Bloch sphere. Beyond these analytically tractable cases, our numerical analysis demonstrates that even highly mixed states can exhibit non-negligible steering probabilities, which increase systematically with the number of measurement settings. Notably, these probabilities exceed those associated with Bell nonlocality by at least one order of magnitude.

Our results further show that ensembles of random states with minimal environmental coupling are typically steerable once the number of settings is sufficiently large. In particular, for $m>2$, more than $50\%$ of sampled states produce steerable correlations, while this fraction exceeds $80\%$ for $m=10$. In the asymptotic limit of infinitely many measurement settings, steering becomes typical in the strongest sense, with the steering probability approaching $100\%$. In contrast, highly mixed-state ensembles can exhibit appreciable steering probabilities, although steering never becomes the dominant feature of such ensembles within the range of parameters considered.

By incorporating random measurements into the steering scenario, we extend previous results on the typicality of Bell nonlocality. The present analysis provides a detailed characterization of the steering measure $\mathcal{P}_S$ across entire ensembles and specialized families of states, such as Bell-diagonal and Werner states. This enables the identification of states and ensembles with the greatest steering potential, highlighting their relevance for protocols where steering serves as a valuable resource.

Several directions remain open for future research. These include extending the analysis to higher-dimensional systems, generalized measurements, and multipartite steering scenarios. Another interesting avenue is the identification of steering inequalities and detection methods that maximize the probability of witnessing steering across broad classes of quantum states.

\section*{Acknowledgements}
GAM and WL are supported by the National Science Centre (NCN, Poland) within the OPUS project (Grant No. 2024/53/B/ST2/04103). This project is supported by the National Research Foundation, Singapore through the National Quantum Office, hosted in A*STAR, under its Centre for Quantum Technologies Funding Initiative (S24Q2d0009). TV acknowledges support from the European Union (CHIST-ERA MoDIC) and from the National Research, Development and Innovation Office NKFIH (Grant Nos.~2023-1.2.1-ERA\_NET-2023-00009 and K145927).

\appendix

\section{Probability of steering a Werner state}
\label{AppA}

Here, we prove that the steering probability of a two-qubit Werner state with visibility $v$ (see Eq.~\eqref{rhoW}), in a two-setting scenario, is given by 
Eq.~\eqref{PSv}. 
To see that, let $A_x=\vec a_x\cdot\vec\sigma$ for $x=1,2$ be the pair of qubit observables with unit Bloch vectors $\|\vec a_x\|=1$. For two settings the assemblage obtained from $\varrho_W(v)$ is steerable if and only if \cite{Uola_2015}
\begin{equation}
\label{Sineq}
S:=\|\vec a_1+\vec a_2\|+\|\vec a_1-\vec a_2\|>\frac{2}{v}.  
\end{equation}
By rotational symmetry we may fix $\vec a_1=(1,0,0)$ and parametrize $\vec a_2$ by the polar angle $\theta\in[0,\pi]$ relative to $\vec a_1$. With this choice we have $\|\vec a_1\pm\vec a_2\|=\sqrt{2(1\pm\cos\theta)}$. So the steering condition~\eqref{Sineq} becomes
\begin{equation}
\label{Stheta}
S(\theta)=\sqrt{2(1+\cos\theta)}+\sqrt{2(1-\cos\theta)}>\frac{2}{v},    
\end{equation}
where $0<v\le 1$. We seek the probability $P_S(v)\equiv\text{Pr}(S>2/v)$ when the measurement directions $\vec a_x$ are chosen independently and uniformly on the Bloch sphere. In this case the induced density for $\theta$ is $p(\theta)=\frac{1}{2}\sin\theta$. Hence
\begin{equation}
\label{PSint}
P_S(v)=\text{Pr}\big(S(\theta)>{2}/{v}\big)
=\frac{1}{2}\int_{\{\theta:\,S(\theta)>2/v\}}\sin\theta\,d\theta.    
\end{equation}
Squaring and rearranging Eq.~\eqref{Stheta} gives us
\begin{equation}
\label{Stv}
\sqrt{1-\cos^{2}\theta}>(1/v^2)-1.   
\end{equation}
For \(v\le 1/\sqrt{2}\) no $\theta$ satisfies the inequality and thus the probability $P_S(v)$ is zero. Assume now $v>1/\sqrt{2}$. Then the right-hand side of Eq.~\eqref{Stv} is nonnegative, so squaring again and rearranging yields
\begin{equation}
\cos^{2}\theta<(2v^{2}-1)/v^{4}.    
\end{equation}
Then the admissible interval in Eq.~\eqref{PSint} is $[\theta^*,\,\pi-\theta^*]$, where we defined $\theta^*=\arccos\!\big(\sqrt{2v^{2}-1}/v^{2}\big)$. Integrating the density $p(\theta)=(1/2)\sin\theta$ over this interval gives
\begin{equation}
\text{Pr}\Big(S>\frac{2}{v}\Big)
= \frac{1}{2}\int_{\theta^*}^{\pi-\theta^*}\sin\theta\,d\theta
=\frac{\sqrt{2v^{2}-1}}{v^{2}}.    
\end{equation}
Combining the two regimes ($v\le 1/\sqrt 2$ and $v> 1/\sqrt 2$) we obtain the piecewise expression for $P_S(v)$ in Eq.~\eqref{PSv}.

\section{Steering probability of the Werner state on the equator}
\label{AppB}

Here, we show that the probability of steering a two-qubit Werner state defined in Eq.~\eqref{rhoW} on the equator is given by Eq.~\eqref{eq:werequator2} for two measurement settings and by Eq.~\eqref{eq:werequator3} for three settings. 
To start, consider $m$ measurement Bloch vectors chosen uniformly on the equator, i.e., with respect to the Haar measure on the angle. In this setting, compact expressions can be derived for the steering probability for the cases of $m=2$ and $m=3$.

The derivation for $m=2$ follows from Appendix~\ref{AppA}. Explicitly, for $v \in [1/\sqrt 2,1]$, the steering probability is given as
\begin{equation}
P_S(v) = \frac{4}{\pi}\arccos\left(\frac{1}{\sqrt{2}\,v}\right).    
\end{equation}
Below the critical threshold of $v=1/\sqrt{2}$, we again have $P_S(v)=0$.

For $m=3$, one can use the fact that there also exists a necessary and sufficient analytic criterion for the steerability of Werner states when the Bloch vectors are restricted to the equator~\cite{Yu_2013}. Using the results from \cite{Yu_2013} and techniques analogous to the ones in Appendix~\ref{AppA} one obtains 
\begin{equation}
P_S(v) = \frac{32}{\pi^2} \int_{0}^{\pi/4} \min\Big(t, \arccos\frac{(1/v) - \cos(2t)}{2 \sin t}\Big) \, dt,   
\end{equation}
where $v \geq 2/3$. For $v < 2/3$, we have $P_S(v)=0$.

\section{Mapping a two-qubit state to a Werner state by local filtering}
\label{AppC}

In this section, we will expand on the argument why the $A \rightarrow B$ steering probability of a Haar random state with product noise shown in Eq.~\eqref{eq:nonsym} follows the one for the Werner state. The state of interest is given as
\begin{align}\label{eq:nonsym3}
\varrho^{AB}(v) =v\ketbra{\psi_H}{\psi_H}+\frac{1-v}{2} \left(\one_2 \ot \tr_A \left(\ketbra{\psi_H}{\psi_H}\right)\right).
\end{align}
For clarity and further use, we express the non-product Haar random state above as
\begin{equation}
\ket{\psi_H} = (U_A\otimes U_B)(\cos\theta\ket{00} + \sin\theta\ket{11}).   
\end{equation}
For statements about the steering probability, restricting to non-product Haar-random states entails no loss of generality, since the set of separable pure states is of zero measure.
Knowing that local filtering on Bob’s side does not change the steerability of the state \cite{Tulio_2015} (fact (i) in the main text) we will apply the local filter on Bob's side
\begin{equation}
\label{eq:FB}
F_B=\lambda\,(-i\sigma_y)\,U_A^*\,D^{-1}U_B^\dagger,   
\end{equation}
where
\begin{equation}
\label{eq:Dmat}
D=\begin{pmatrix}
\cos\theta & 0\\
0 & \sin\theta
\end{pmatrix},    
\end{equation}
with $0<\theta<\pi/2$ and  $\lambda>0$ is chosen small enough so that $F_B^{\dagger}F_B\le \one_2$. 
With this choice of $D$ and the $F_B$ in Eqs.~\eqref{eq:Dmat},~\eqref{eq:FB}, the filtered pure state becomes
\begin{equation}
(\one_2\otimes F_B)\ket{\psi_H}
=
\lambda\,(U_A\otimes (-i\sigma_y)U_A^*)
\left(\ket{00}+\ket{11}\right).
\end{equation}
Now, since
$(U_A\otimes U_A^*)\ket{\Phi^+}=\ket{\Phi^+}$, it follows that 
\begin{equation}
(\one_2\otimes F_B)\ket{\psi_H} = \sqrt{2}\lambda\ket{\psi^-},
\end{equation}
so the state $\ket{\psi_H}$ is mapped to the singlet state.

At the same time, the reduced state of Bob transforms as
\begin{equation}
F_B\left(\tr_B \left(\ketbra{\psi_H}{\psi_H}\right)\right)F_B^{\dagger} = \lambda^2\one_2.    
\end{equation} Putting the two parts together, the filtered two-qubit state is 
\begin{equation}
(\one_2\otimes F_B)\rho^{AB}(v)(\one_2\otimes F_B)=2\lambda^2(v\ketbra{\psi^-}{\psi^-}+(1-v)\one_4/4).    
\end{equation}
Therefore, after normalization, the resulting state is exactly the Werner state $\rho_W(v)$ and steering probabilities $A \rightarrow B$ are thus equal.

\bibliographystyle{apsrev4-2}
\bibliography{re}

\end{document}